# A Wireless Reconfigurable Metasurface for Enhanced Parallel Magnetic Resonance Imaging

Yuhan Liu[1,2†], Xia Zhu[1,2†], Ke Wu[1,2], Artem Kaliaev[3], Christina A. LeBedis[3], Stephan W. Anderson[2,3*], and Xin Zhang[1,2,4,5,6*]

[1]Department of Mechanical Engineering, Boston University, Boston, Massachusetts, USA

[2]Photonics Center, Boston University, Boston, Massachusetts, USA

[3]Chobanian & Avedisian School of Medicine, Boston University Medical Campus, Boston, Massachusetts, USA

[4]Department of Electrical & Computer Engineering, Boston University, Boston, Massachusetts, USA

[5]Department of Biomedical Engineering, Boston University, Boston, Massachusetts, USA

[6]Division of Materials Science & Engineering, Boston University, Boston, Massachusetts, USA

[†]These authors contributed equally to this work

*Corresponding author: Xin Zhang (xinz@bu.edu); Stephan W. Anderson (sande@bu.edu);

Contributing authors: yuhanyh@bu.edu; xiaz@bu.edu;

wk0305ok@bu.edu; akaliaev@bu.edu; clebedis@bu.edu

## Abstract

Modern magnetic resonance imaging (MRI) relies on application-specific multi-channel receive coils to achieve high performance, but these coils are typically costly, rigid, and difficult to generalize across anatomies. Recent wireless, low-cost metamaterials offer improved signal-to-noise ratio (SNR) but remain anatomy-dependent, are prone to destructive inter-element interference, and lack demonstrated compatibility with parallel imaging. Herein, a wireless, reconfigurable coaxial loop metasurface (CLM) is introduced as a platform for localized SNR enhancement that can operate either as a standalone element or as an insertable add-on alongside existing clinical receive systems. Through its coaxial architecture and shared current pathways, the CLM establishes a collective in-phase resonant mode that enforces phase-coherent current distributions across all loops, resulting in consistently constructive interference. Benchmarking on a 3.0 T MR system using an 8-loop CLM shows SNR enhancements of up to 14.8-fold and 14.02-fold in the sagittal and axial planes, relative to the birdcage coil (BC). As an add-on to a clinical posterior receive array, it further demonstrates up to 2.9-fold SNR enhancement and compatibility with parallel imaging across ex vivo and in vivo settings. The proposed CLM paves the way toward a new class of reconfigurable and insertable MRI hardware for flexible and system-compatible signal enhancement.

## Introduction

Magnetic resonance imaging (MRI) stands as one of the most powerful and widely used modalities in clinical diagnostics, offering superior contrast between soft tissues and bones without the risks associated with ionizing radiation [1,2]. The foundation of MRI lies in the nuclear magnetic response of atomic nuclei. When exposed to a strong static magnetic field ($B_0$), nuclear spins partially align, creating a net magnetization along the field direction. Application of an alternating radiofrequency (RF) excitation field ($B_1^+$) tips these nuclei away from equilibrium; as they subsequently relax back toward their original alignment, they emit detectable signals in the form of the $B_1^-$ field [3,4]. These signals encode information about anatomical structures, tissue composition, and pathological changes. In practice, the diagnostic quality of MRI is largely determined by the signal-to-noise ratio (SNR) of the acquired images [5-8].

To maximize SNR, modern MRI systems have adopted high-density, application-specific multi-channel receive arrays as the gold standard, enhancing RF reception in targeted regions of interest (ROIs) and delivering high-quality anatomical imaging [9,10]. However, this approach introduces significant practical and economic burdens. The rigid and highly specialized nature of multi-channel coils limits their versatility, as each coil is typically designed for a specific body part such as the head [11-13], breast [14,15], knee [16], or torso [17]. Hospitals are consequently forced to maintain costly inventories of coil sets to accommodate diverse clinical applications, and the setup of these wired coils, including patient positioning and cable management, is often time-consuming and cumbersome [18]. There is therefore an urgent need for new technologies that provide substantial SNR enhancement while overcoming the high-cost, rigidly specialized limitations of current coil systems.

Rather than replacing existing coil systems, an ideal solution is to augment their performance through compact, insertable modules that enhance local sensitivity while leveraging existing hardware. Even with a comprehensive coil inventory, many anatomical regions characterized by complex or confined geometries remain inadequately served by dedicated coil designs [19,20]. In current clinical practice, imaging of such regions often relies on large-coverage, general-purpose receive arrays, such as spine or torso coils, which provide broad but suboptimal sensitivity in the targeted area [21,22]. Importantly, current MRI systems lack a practical mechanism to locally augment receive sensitivity on top of existing coil infrastructures without hardware replacement. This limitation cannot be effectively addressed by simply expanding the coil portfolio; instead, it calls for a compact, insertable, and reconfigurable enhancement device that can be deployed in proximity to the ROI and adapted to diverse anatomical scenarios. Furthermore, modern clinical workflows rely heavily on parallel imaging techniques to accelerate acquisition, making compatibility with multi-channel receive arrays a critical requirement for clinical translation [23-25]. Any such supplementary device must therefore not only provide localized SNR enhancement, but also remain compatible with the spatial encoding framework of multi-channel receive systems that underpins accelerated imaging.

In recent years, wireless metamaterials have emerged as a promising strategy to overcome the limitations of conventional coil systems, owing to their subwavelength resonance properties that enable compact and efficient designs. Early designs, including conducting Swiss rolls [26], capacitively loaded ring resonators [27,28], helical coil [29,30], and parallel-wire resonators [31-33], demonstrated the ability to redistribute the $B_1^-$ field in the near-field region, thereby enhancing local signal sensitivity and improving SNR. These devices are typically lightweight, low-cost, and fabricated using printed circuit boards (PCBs) or conductive materials. However, many of these early structures lack full compatibility with modern clinical MRI systems and standard imaging sequences. Their primary limitation lies in the inability to selectively respond to the RF field during different phases of the MRI cycle. Without appropriate detuning mechanisms, these passive metamaterials may interact undesirably with the transmit $B_1^+$ field, leading to field distortion and potential degradation of image uniformity and patient safety [34]. Consequently, despite their initial promise, these designs have seen limited practical adoption in clinical MRI environments.

More recently, metamaterials with improved clinical compatibility have been developed through the incorporation of self-detuning mechanisms based on PIN diodes or varactors [35-38], and therefore deactivate the structure during the transmit phase, ensuring safety and preserving $B_1^+$ field uniformity. Nonetheless, to achieve broader spatial coverage, multiple metamaterial unit cells are often arranged in arrays, which introduces a fundamental physical limitation: inter-element coupling. This coupling induces opposing currents on adjacent conductors among neighboring unit cells [39-41], leading to destructive interference and localized low signal regions within the ROI. As the array density increases, these effects become more pronounced, constraining both scalability and flexibility. Moreover, the majority of existing metamaterial designs have been demonstrated exclusively in conjunction with the scanner's built-in body coil (BC), and their compatibility with multi-channel receive arrays and parallel imaging acceleration remains largely unexplored [42-45].

In this work, a reconfigurable coaxial loop metasurface (CLM) is developed to address the limitations of both conventional coil arrays and existing metamaterial designs (**Figure 1a**). The CLM exploits the inherent shielding of coaxial cables and an engineered PCB architecture that provides a shared current path between adjacent rungs, establishing a collective electromagnetic coupling that maintains all loops nested and their currents in phase. This intrinsic synchronization produces constructive interference across the arrayed loops and eliminates the dark regions commonly observed in traditional metamaterial assemblies, while strategically positioned non-magnetic PIN diodes ensure a selective RF response as passively detuned during transmission. This behavior is further illustrated in **Figure 1b**, where two array configurations with identical dimensions and unit cell numbers are compared through their simulated magnetic field distributions. The in-phase configuration, as realized in the CLM, produces a broad and continuous field profile due to constructive interference, whereas out-of-phase arrays exhibit pronounced field minima between adjacent elements due to destructive interference, with low-signal regions inevitably forming between neighboring unit cells. Leveraging this physical stability, the CLM enables on-demand customization, enabling clinicians to freely add or remove loop elements to tailor the enhancement area according to patient-specific diagnostic requirements. We systematically characterize the electromagnetic evolution of such arrayed rung structures, providing a theoretical guideline for their scalable deployment. Moreover, the thin, planar form factor of the CLM allows it to function as a passive, insertable add-on that can be freely placed alongside or between standard clinical receive arrays in various orientations without requiring any electrical connection, hardware modification, or alteration to scan protocols, providing immediate and localized SNR enhancement in geometrically complex anatomical regions where dedicated coils are unavailable. Importantly, the CLM operates as a wireless, passive device; it preserves the spatial encoding diversity among the surrounding receive channels, maintaining full compatibility with parallel imaging acceleration. Bench measurements and MRI experiments conducted on a 3.0 T clinical scanner (Philips Healthcare) validate the CLM's performance across standalone enhancement with the BC, versatile integration with a clinical posterior receive array, and parallel imaging acceleration factors up to 4. Overall, the proposed CLM establishes a new paradigm in MRI signal enhancement, transforming the technology from an expensive, rigid, and application-specific solution into a low-cost, insertable, and universally compatible platform that seamlessly integrates into diverse clinical workflows.

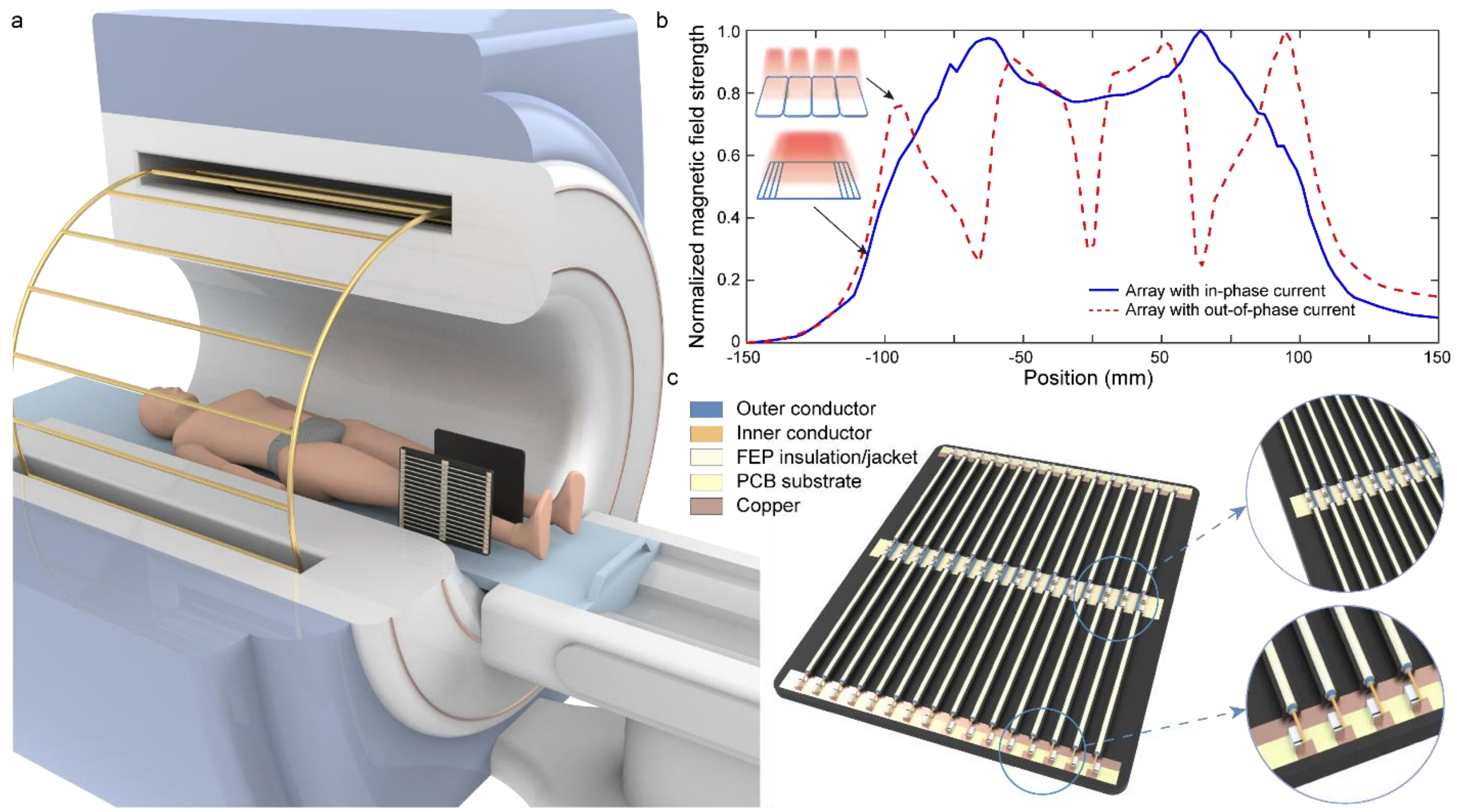

**Figure 1| Coaxial loop metasurface insert for wireless enhancement of MRI performance.** a) Schematic illustration of the MRI setup, in which two CLM inserts are positioned vertically on the patient table and conformally placed around the calf to enhance the receive sensitivity. b) Simulated magnetic field distributions for a 4-loop CLM and a 4-unit-cell conventional array with identical dimensions (274 mm × 168 mm). The in-phase configuration (CLM) produces a broad and continuous field profile, whereas the out-of-phase configuration shows localized signal depressions between neighboring elements. c) Structural details of an 8-loop CLM, comprising 16 coaxial rungs interconnected through shared PCB pathways on both sides. Each coaxial rung incorporates four PIN diodes that bridge the inner and outer conductors, enabling passive, power-

dependent detuning during transmission. The MRI scanner, patient model, and metamaterial models were created in SolidWorks and rendered in KeyShot.

## Results

### 2.1 Design and Principle of the CLM

The CLM is designed as a compact and modular structure for MRI signal enhancement. It consists of 16 parallel coaxial cable rungs connected by three printed circuit boards, with overall dimensions of 230 mm in length and 130 mm in width (**Figure 1c**). Each coaxial rung has an outer diameter of 2.54 mm and comprises four concentric layers: 1) a silver-plated copper inner conductor, 2) a fluorinated ethylene propylene (FEP) dielectric, 3) a silver-plated copper braided outer conductor, and 4) an FEP protective jacket. Each rung incorporates PIN diode pairs that are oriented in opposite directions, and its outer conductor is electrically connected at both ends through the PCBs. At resonance, these rungs are naturally divided into two groups on opposite sides of the metasurface, carrying currents in opposite directions. These rungs then formed closed loops through the shared PCB paths, enabling controlled current distribution and tailored magnetic field profiles. The PCB layout preserves electrical continuity in the outer conductor at both ends while intentionally interrupting the inner conductor at the center. This configuration ensures that currents induced in all loops remain in phase, leading to inherently constructive interference across the metasurface. As a result, adding or removing a symmetric pair of rungs at equal distances from the center automatically creates or removes an additional resonant loop that constructively couples with existing loops, without inducing out-of-phase currents between adjacent conductive elements as observed in conventional arrayed structures (**Figure S1**). This intrinsic property enables a mechanically robust and reconfigurable structure in which loops can be added or removed while maintaining a continuous and constructive field distribution, thereby the formation of low-signal regions.

### 2.2 Unit Cell Architecture and Electromagnetic Behavior

Wireless metamaterial structures used in MRI are designed to interact with RF electromagnetic fields, and their effectiveness depends critically on how selectively they couple to different components of the RF excitation. While designed primarily to enhance the magnetic field, conventional resonant conductors operating at the Larmor frequency inevitably generate electric field components as well, since electromagnetic resonance necessarily involves energy exchange between the two. However, because these structures are not engineered to manage the electric field, it remains weakly confined and extend freely into the surrounding space. When such structures are placed in proximity to an imaging subject, stray electric fields can penetrate biological tissue, making the resonance highly sensitive to environmental loading and leading to frequency shifts, degradation of imaging performance, and potential safety concerns, as electric fields are the dominant contributor to specific absorption rate (SAR). The coaxial architecture of the CLM addresses these limitations by promoting strong electric-field confinement. At resonance, the large structural capacitance between the inner and outer conductors localizes the electric field primarily within the coaxial layers, substantially suppressing stray electric fields and reducing capacitive coupling to the imaging subject (**Figure S2**). This confinement also enhances operational robustness by minimizing environmental perturbations to the resonance frequency. Because the electromagnetic response is primarily governed by the capacitance between the inner and outer conductors, rather than parasitic capacitance between the CLM and the imaging subject, the resulting resonance preferentially contributes to magnetic field enhancement while limiting unwanted electric field exposure.

In addition to electric field confinement, the CLM must selectively interact with the alternating RF field during MRI, resonating during signal reception to enhance the weak $B_1^-$ field while remaining transparent during high-power $B_1^+$ transmission to avoid SAR elevation and excitation profile distortion. This selective response is achieved through a self-detuning mechanism that leverages the unit cell architecture of the CLM.

The electromagnetic behavior of the CLM can be understood by examining the structure of each coaxial rung in detail. As illustrated in **Figure 2a**, each rung can be decomposed into two unit cells, each containing a central gap in the outer conductor and a pair of PIN diodes bridging the inner and outer conductors. The two PIN diodes within each unit cell are oriented in opposite directions, ensuring that during each half of the RF cycle, at least one diode can be forward-biased to short-circuit the inner and outer conductors. At the boundaries between adjacent unit cells, the outer conductors maintain electrical continuity through the shared PCB paths, while the inner conductor is intentionally interrupted, forming an effective inner gap. Each resonant loop of the CLM thus comprises two rungs, equivalently four unit cells, closed at both ends through the shared PCB paths.

**Figure 2b** presents the equivalent circuit model of the CLM and illustrates the nested loop topology that governs its collective resonance. Within each unit cell, the outer conductor gap introduces a gap capacitance ($C_g$), while the structural capacitance between the inner and outer conductors on each side of the gap is represented by $C_s$. Each unit cell also possesses a self-inductance ($L_0$) and resistance ($R_0$). When multiple rungs are assembled, the resulting loops are concentrically nested, sharing substantial mutual magnetic flux. This geometric overlap gives rise to strong mutual inductive coupling ($k_{12}$, $k_{13}$, ...) between all loop pairs, which is the physical origin of the in-phase, constructive current distribution in the CLM. The equivalent circuit further clarifies the self-detuning mechanism. During the reception phase, the RF power is insufficient to forward-bias the PIN diodes; the diodes behave as open circuits, and the large structural capacitance $C_s$ fully participates in resonance, tuning the CLM to the Larmor frequency for effective signal enhancement. During the transmission phase, the high-power $B_1^+$ field forward-biases the diodes, which act as closed switches that create a conductive path between the inner and outer conductors and short-circuit $C_s$. With this dominant capacitance bypassed, the resonance condition is no longer satisfied at the Larmor frequency, and the CLM is effectively detuned, preventing unwanted interaction with the transmit field.

This nonlinear, power-dependent response is experimentally validated through reflection coefficient measurements using a VNA. **Figure 2c** shows the measured reflection spectra as the excitation power increases from -12 dBm to 10 dBm, revealing a progressive downward shift in resonance frequency accompanied by a reduction in resonance strength. **Figure 2d** compares the reflection coefficients measured with and without PIN diodes over a broader excitation power range (−25 dBm to 15 dBm). With PIN diodes incorporated, the CLM exhibits a strong resonance at low excitation powers, corresponding to the reception phase. As the excitation power increases, the resonance is progressively suppressed, in contrast to the power-independent behavior observed without PIN diodes. In practical MRI operation, transmit power levels are orders of magnitude higher than those used in these bench measurements, ensuring complete detuning during transmission. This self-adaptive behavior allows the CLM to be safely integrated into MRI systems without requiring modifications to existing hardware or pulse sequences.

When resonant during signal reception, the current distribution within each unit cell is governed by the skin effect and the engineered coaxial geometry. Under RF excitation, currents are confined to the conductor surfaces and give rise to three distinct components: the current on the inner conductor ($I_i$), the current on the inner surface of the outer conductor ($I_{oi}$), and the current on the outer surface of the outer conductor ($I_{oo}$), as illustrated in **Figure 2e**. The induced currents are primarily established on the outer surface of the outer conductor. At the outer conductor gaps, charge accumulation produces a voltage that drives current along the inner surface of the outer conductor; through inductive coupling, a mirrored current is simultaneously induced on the inner conductor, flowing in the opposite direction. The simulated current distributions along the rung length (**Figure 2f**) show that $I_i$ exhibits a symmetric profile, peaking near the outer gaps and decreasing to zero at the central inner gap and at both ends. $I_{oi}$ follows the same spatial profile but in the opposite direction with equal magnitude, resulting in mutual cancellation of their magnetic field contributions. In contrast, $I_{oo}$ remains nearly uniform along the rung and serves as the primary source of magnetic field enhancement. This spatial uniformity of $I_{oo}$ produces a homogeneous amplified magnetic field distribution across the metasurface.

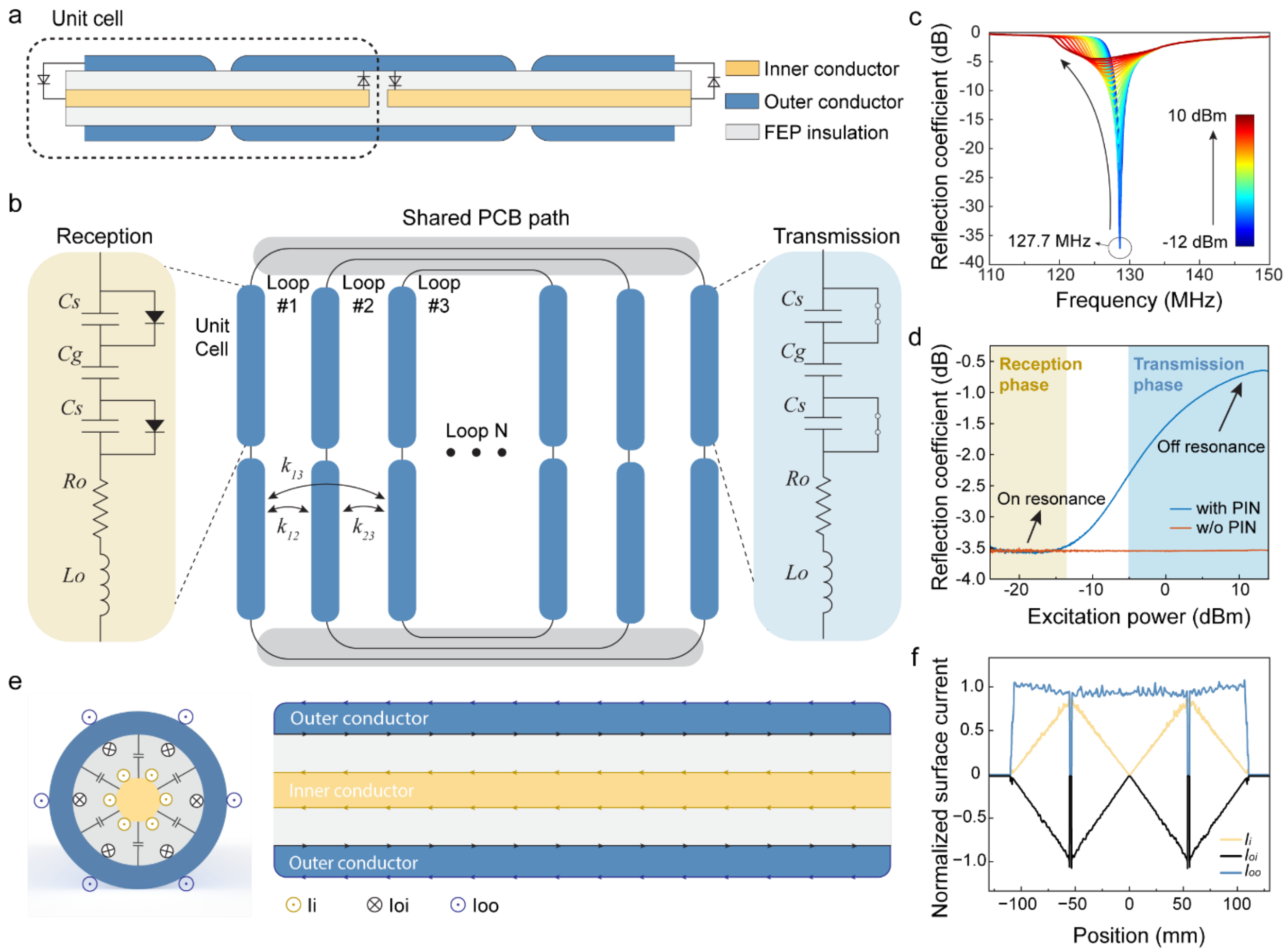

**Figure 2| Electromagnetic characterization and working principle of the CLM.** a) Structural decomposition of a coaxial rung into two unit cells. b) Equivalent circuit and collective topology of the CLM. The reception and transmission states are illustrated, highlighting passive self-detuning via PIN diodes under high RF power. c) Measured reflection coefficient ($S_{11}$) spectra as a function of excitation power (-12 to 10 dBm), showing progressive suppression and frequency shift of the resonance with increasing power. d) Reflection coefficient as a function of excitation power with and without PIN diodes at 127.7 MHz. Insets show representative simulated field distributions in receiving and transmitting states. e) Cross-sectional view of the coaxial structure and corresponding current pathways, illustrating the three current components on the inner

conductor ($I_i$), the inner surface of the outer conductor ($I_{oi}$), and the outer surface of the outer conductor ($I_{oo}$). f) Simulated current distributions along each coaxial rung.

### 2.3 Uniform and Reconfigurable Enhancement from Collective Behavior

While the current distribution and nonlinear response are characterized at the level of individual coaxial rungs, the distinctive performance of the CLM emerges from the collective electromagnetic behavior of multiple rungs assembled into an array. To elucidate this collective response, full-wave electromagnetic simulations were performed for CLM configurations comprising 4, 8, 12, and 16 rungs. **Figure 3a** illustrates the induced current distribution when the CLM is tuned to the Larmor frequency. In this resonant state, the N coaxial rungs are naturally organized into N/2 nested loops. Rungs on each half of the metasurface carry currents flowing in the same direction, and the two halves are electrically connected through shared PCB paths to form closed loops. Unlike conventional multi-loop structures in which neighboring elements often support opposing or out-of-phase currents on adjacent conductors, all loops in the CLM resonate in phase. This in-phase current distribution leads to strong constructive interference, producing an enhanced magnetic field oriented perpendicular to the metasurface plane. This coherent behavior constitutes the fundamental mechanism enabling the CLM to reconfigure and tailor magnetic field enhancement without introducing destructive interference.

The distribution of current amplitude across the rungs further reflects this collective behavior. **Figure 3b** shows the simulated outer-surface current amplitude ($I_{oo}$) and its fitted curve at each rung position in an 8-loop CLM. Each data point represents the current magnitude carried by an individual rung, and the fitted curve across all 16 points follows a near-sinusoidal envelope: the central rungs carry the largest currents, with the amplitude decreasing symmetrically toward the outer rungs. This spatial profile is a direct consequence of the nested loop topology of the CLM, in which inner loops enclose a smaller area and therefore support a higher current density, while progressively larger outer loops carry proportionally lower currents. Importantly, this graded amplitude distribution coexists with the uniform phase across all rungs, and the two together govern the overall field profile of the CLM: the in-phase currents ensure constructive interference and eliminate signal voids, while the amplitude envelope naturally concentrates the peak magnetic field enhancement toward the center of the metasurface, yielding a spatially well-defined and homogeneous enhancement region within the ROI.

The impact of this coherent current distribution on the magnetic field is illustrated in **Figure 3c**, which shows simulated magnetic field maps in the sagittal and axial planes. For the 2-loop configuration, the magnetic field is strongly localized near the conductors, leaving a pronounced low-field region at the center. As the number of loops increases, the magnetic flux generated by individual loops coherently superposes, shimming the central void and redistributing the field across the metasurface. The 8-loop configuration exhibits a continuous, uniform high-field region spanning the entire metasurface area, in sharp contrast to conventional metamaterial arrays where destructive interference manifests as pronounced dark regions between adjacent unit cells [37-39].

**Figure S3** provides a quantitative comparison of magnetic field strength along the dashed lines indicated in **Figure 3c**. In the sagittal plane, increasing the number of loops does not merely redistribute the field toward the center of the ROI; instead, it leads to an overall enhancement of magnetic field strength. Specifically, the 8-loop CLM produces an approximately 4.5-fold increase in magnetic field amplitude compared with the 2-loop configuration. In the axial plane, all configurations show a peak magnetic field near the metasurface that decays with distance. Among these configurations, the 8-loop CLM achieves the highest magnetic field strength near the metasurface, while at larger distances its performance becomes comparable to that of the 6-loop configuration, reflecting reduced sensitivity to further loop densification along the direction normal to the metasurface. Further insight is provided by the field energy analysis in **Figure 3d**, evaluated at the central points $P_1(0,0,0)$ and $P_2(0,0,15)$. As the number of loops increases from 2 to 8, the magnetic field energy rises significantly while the electric field energy decreases concurrently. This inverse trend indicates improved field confinement with increasing loop count: electric fields are progressively concentrated within the metasurface structure, while magnetic fields are more efficiently projected into the imaging volume. This characteristic is critical for maximizing SNR, as it reduces electric-field-induced sample noise and mitigates the risk of localized SAR hotspots.

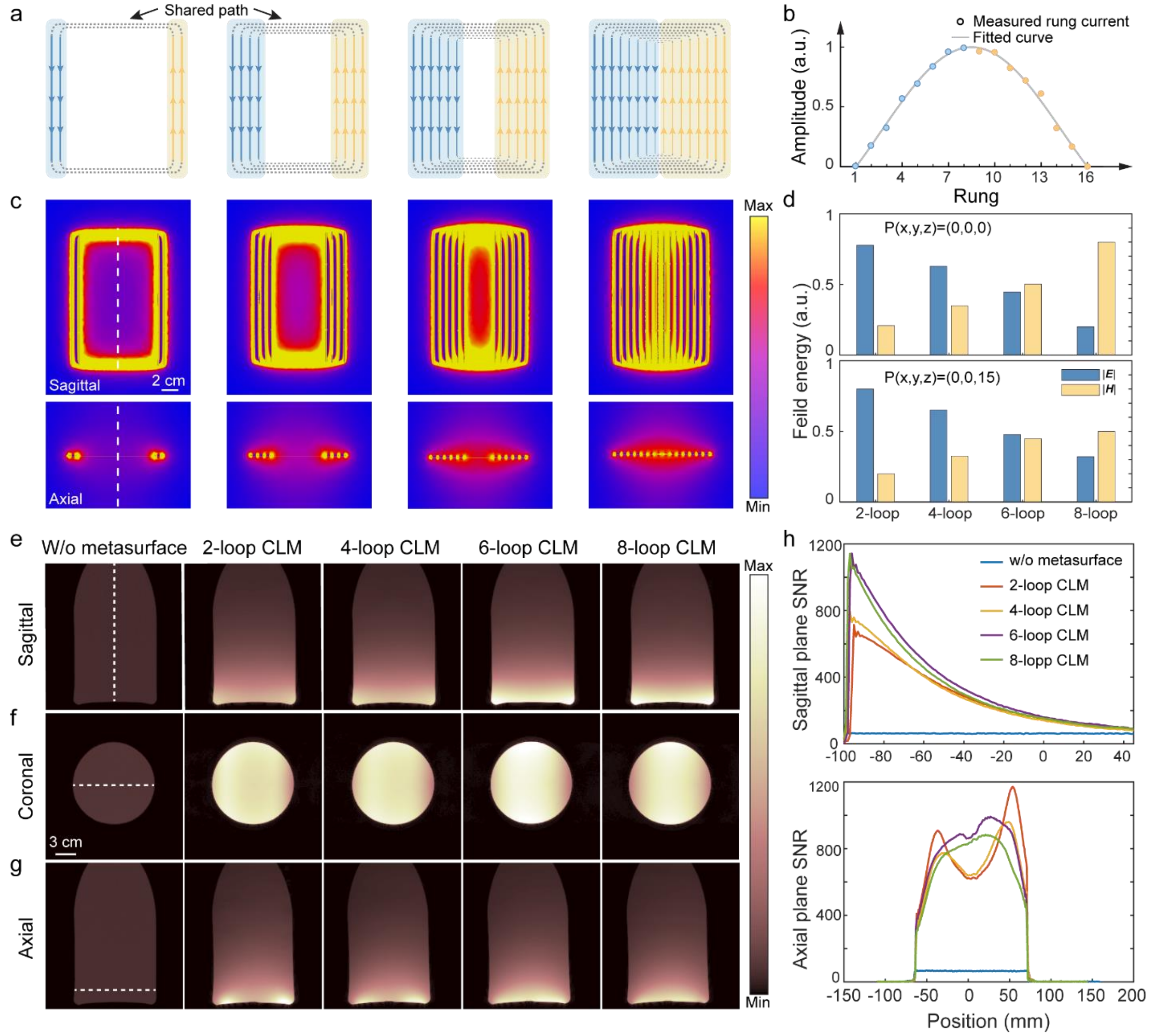

**Figure 3 | Current distribution–driven reconfigurability and MRI validation of the CLM.** a) Schematic illustration of the collective current distribution in CLMs with 4, 8, 12 and 16 rungs. b) Simulated current amplitude of the outer-surface current ($I_{oo}$) across the rungs for an 8-loop CLM. c) Simulated magnetic field distributions at 127.7 MHz for CLMs with 2, 4, 6, and 8 loops. Increasing the number of loops progressively redistributes the magnetic field from edge-localized regions toward the center, resulting in a more uniform and continuous high-field region. d) Simulated electric and magnetic field energies evaluated at the center point (P = (0, 0, 0)) and 15

mm above the metasurface (P = (0, 0, 15)). e-g) Phantom MRI results acquired without the metasurface and with 2-, 4-, 6-, and 8-loop CLMs in the sagittal (e), coronal (f), and axial (g) planes. h) SNR profiles extracted along the dashed lines in (e) and (g). Scale bars in (e) to (g) are 3cm.

**2.4 MRI validation with a Mineral Oil Phantom**

To experimentally validate the constructive interference mechanism of the CLM under realistic MRI conditions and to quantify its SNR enhancement, phantom imaging experiments were performed on a clinical 3.0 T MRI system (Philips Healthcare). A cylindrical mineral oil phantom with a diameter of 160 mm was positioned directly above the CLM and centered at the isocenter of the BC, which was used for both RF transmission and reception. **Figures 3e-g** show representative MR images acquired in the sagittal, coronal, and axial planes without the CLM and with CLM configurations consisting of 2, 4, 6, and 8 loops. The sagittal imaging plane is parallel to the coaxial rungs and exhibits uniform signal enhancement across the phantom. In contrast, the axial plane is perpendicular to the rungs, and the corresponding images clearly reveal the loop-dependent field distribution, providing direct visual evidence of constructive superposition among individual loops. Coronal images in **Figure 3f** are acquired at a height of 10 mm above the phantom base. Across all three imaging planes, increasing the number of loops does not introduce dark regions between adjacent loops in the SNR maps, confirming the nested, in-phase current distribution that underpins the reconfigurable behavior of the CLM.

Quantitative comparisons of SNR profiles extracted along the dashed lines in each imaging plane are shown in **Figure 3h** and **Figure S4**. In the sagittal plane, the 2-loop CLM yields an approximately 8.1-fold SNR enhancement compared with the BC alone. As the number of loops increases, the peak SNR rises from approximately 630 for the 2-loop configuration to about 1160 for the 6-loop configuration, after which it saturates, with similar peak values observed for the 6- and 8-loop CLMs. Overall, the 8-loop CLM achieves a maximum SNR enhancement of approximately 14.8-fold relative to imaging without the CLM. In the axial plane, the 2-loop CLM produces an SNR profile with two distinct peaks whose separation corresponds to the inner loop width, directly reflecting the CLM's spatially localized enhancement. As additional loops are incorporated, these two peaks progressively converge toward the center. In the 8-loop configuration, the peaks merge into a single central maximum as the effective inner loop width reaches its minimum. Notably, in the axial plane, increasing the number of loops primarily reshapes and localizes the enhanced magnetic field region while maintaining comparable overall SNR enhancement, with an 8-loop CLM achieving peak enhancements of approximately 14.02-fold. This behavior indicates that the CLM primarily redistributes the magnetic field while maintaining a strong overall enhancement. The axial SNR maps visually capture this trend, showing that signal enhancement progressively converges from both sides toward the center as additional loops are introduced, leading to increased signal concentration at the core of the ROI. As a result, the SNR enhancement becomes more spatially uniform across the entire ROI. These observations confirm that the CLM can flexibly redistribute the magnetic field without compromising overall signal enhancement, thereby demonstrating its reconfigurability.

A comparison between the CLM and other metasurface designs employing parallel conductive rungs or arrayed resonant structures is summarized in **Table S1**, Supporting Information. Benefiting from the intrinsic electric-field shielding of the coaxial architecture and the shared PCB-based circuit design, the CLM exhibits self-detuning capability while maintaining excellent reconfigurability and tunability without compromising SNR. This approach effectively avoids the destructive interference that commonly arises when scaling conventional metamaterial arrays. Within a practical tuning range, the CLM therefore supports application-specific and personalized adjustments to meet diverse ROI imaging requirements.

### 2.5 Versatile Integration with a Clinical Posterior Receive Array

The preceding experiments validate the CLM's enhancement capability using the BC for both transmission and reception. While this configuration isolates the intrinsic electromagnetic behavior of the CLM, clinical MRI workflows predominantly employ multi-channel receive arrays to achieve the spatial encoding diversity required for parallel imaging acceleration. To evaluate the CLM's compatibility and versatility as an insertable add-on within such clinical configurations, a series of phantom experiments were conducted using a dStream FlexCoverage Posterior coil (Philips Healthcare) — a multi-channel spine array integrated into the MRI patient table — as the primary receive element. The BC was used exclusively for RF transmission in all configurations.

Four representative setups were designed to systematically assess the CLM's performance in different spatial orientations relative to the receive array (**Figure 4a** and **Figure S5**). In Setup #1 (reference), the cylindrical mineral oil phantom was placed directly on the posterior coil without any CLM, establishing the baseline SNR distribution. In Setup #2, a single CLM was positioned vertically alongside the phantom, perpendicular to the posterior coil surface. In Setup #3, the CLM was placed horizontally on top of the phantom, parallel to and directly opposing the posterior coil, forming a sandwich-like configuration. In Setup #4, two CLMs were positioned bilaterally on opposite sides of the phantom, both perpendicular to the posterior coil, creating a three-sided enhancement geometry. These configurations represent canonical spatial arrangements that collectively demonstrate the orientational flexibility of the CLM as an insertable device.

**Figures 4a-d** present the SNR maps and corresponding line profiles for all four setups across the axial, sagittal, and coronal planes. In the reference configuration (Setup #1), the SNR distribution is strongly asymmetric, with signal intensity concentrated near the posterior coil and decaying rapidly with distance—a characteristic limitation of single-sided surface arrays. Upon introduction of the CLM, each configuration produces directional SNR enhancement that is spatially correlated with the metasurface position and orientation. Setup #3, in which the CLM opposes the posterior coil in a sandwich geometry, yields the most pronounced enhancement in the sagittal plane, with peak SNR reaching approximately 1200—roughly 3-fold higher than the reference—as the combined sensitivities of the two devices converge within the intervening volume. Setup #2 provides strong lateral enhancement with a peak SNR of approximately 600 in the axial plane, while Setup #4 extends this to a symmetric bilateral profile. Across all configurations, the CLM enhances SNR in its respective direction while maintaining the baseline performance of the posterior coil without introducing imaging artifacts.

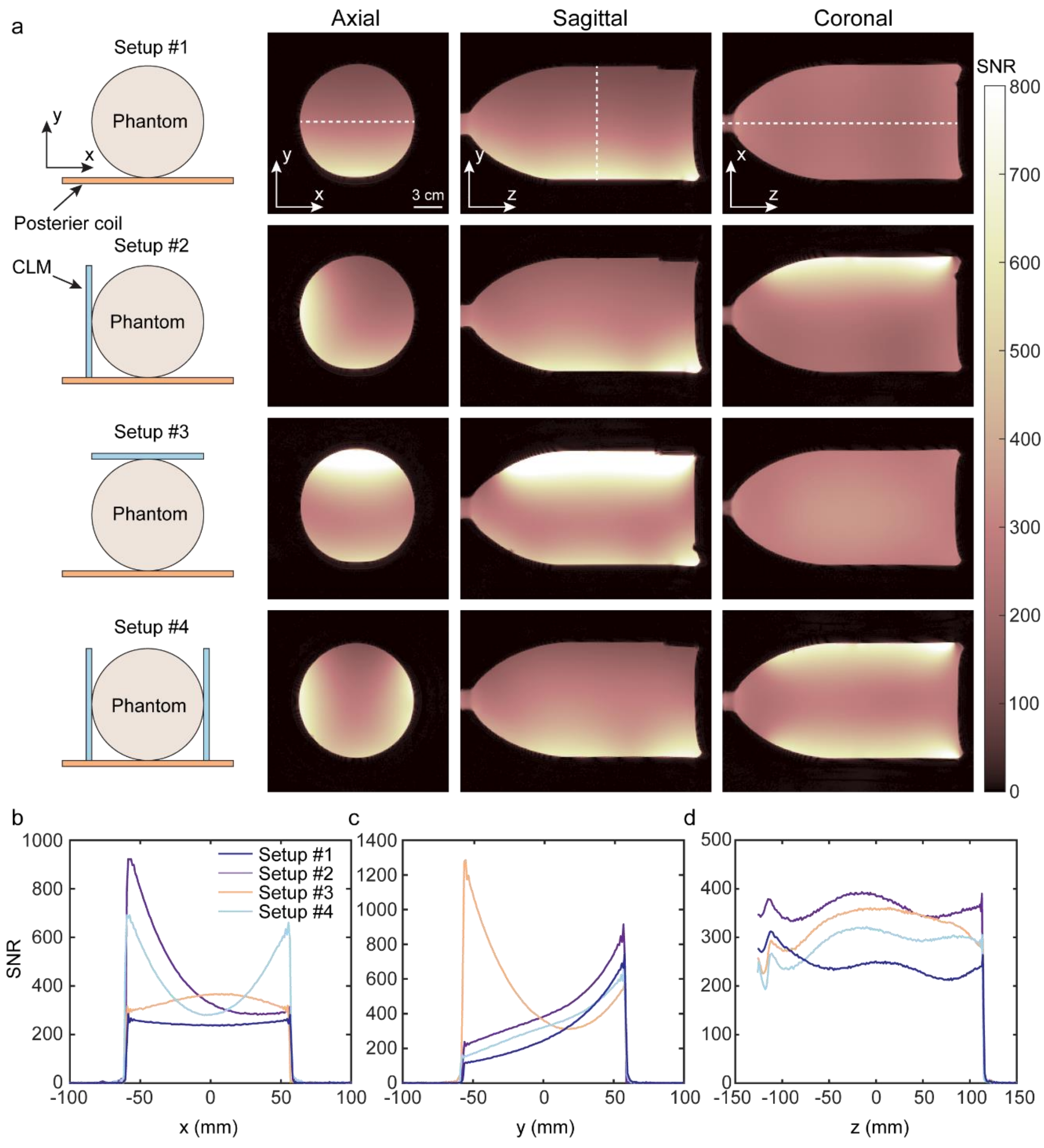

**Figure 4. Integration of the CLM with a clinical multi-channel receive array.** a) Four experimental configurations illustrating the spatial arrangement of the CLM relative to a posterior receive array. Corresponding SNR maps in the axial, sagittal, and coronal planes are shown for each setup. b-d) SNR line profiles extracted along the dashed lines in (a) for the axial (b), sagittal (c), and coronal (d) planes. The introduction of the CLM redistributed the SNR enhancement

moved from near the posterior coil to where the metasurfaces are placed. Across all configurations, the CLM enhances SNR without degrading the baseline performance of the receive array. Scale bars in (a) are 3cm.

To further assess whether the CLM remains compatible with parallel imaging acceleration, Setup #4 (bilateral CLMs with posterior coil) and Setup #1 (reference) were compared under SENSE acceleration factors of P = 2, 3, and 4. **Figure 5a** shows coronal SNR maps for both configurations at each acceleration factor. As expected, increasing acceleration reduces the overall SNR for both setups due to the combined effects of reduced sampling and geometry-factor-related noise amplification. However, the SNR enhancement ratio between Setup #4 and Setup #1, shown in **Figure 5b and c**, indicates that the enhancement remains consistent across all acceleration factors, exhibiting a spatially uniform gain with only minimal variation across different acceleration factors. This consistency suggests that, while the CLM increases the sensitivity magnitude of the surrounding receive channels—which underlies the observed SNR enhancement—it does not substantially alter the spatial encoding patterns among channels required for parallel imaging reconstruction. The passive and wireless nature of the CLM further allows it to be incorporated without direct electrical interaction with the receive array, and the maintained enhancement across acceleration factors is consistent with its compatibility with parallel imaging workflows, without evidence of degradation in reconstruction performance or the need for additional calibration.

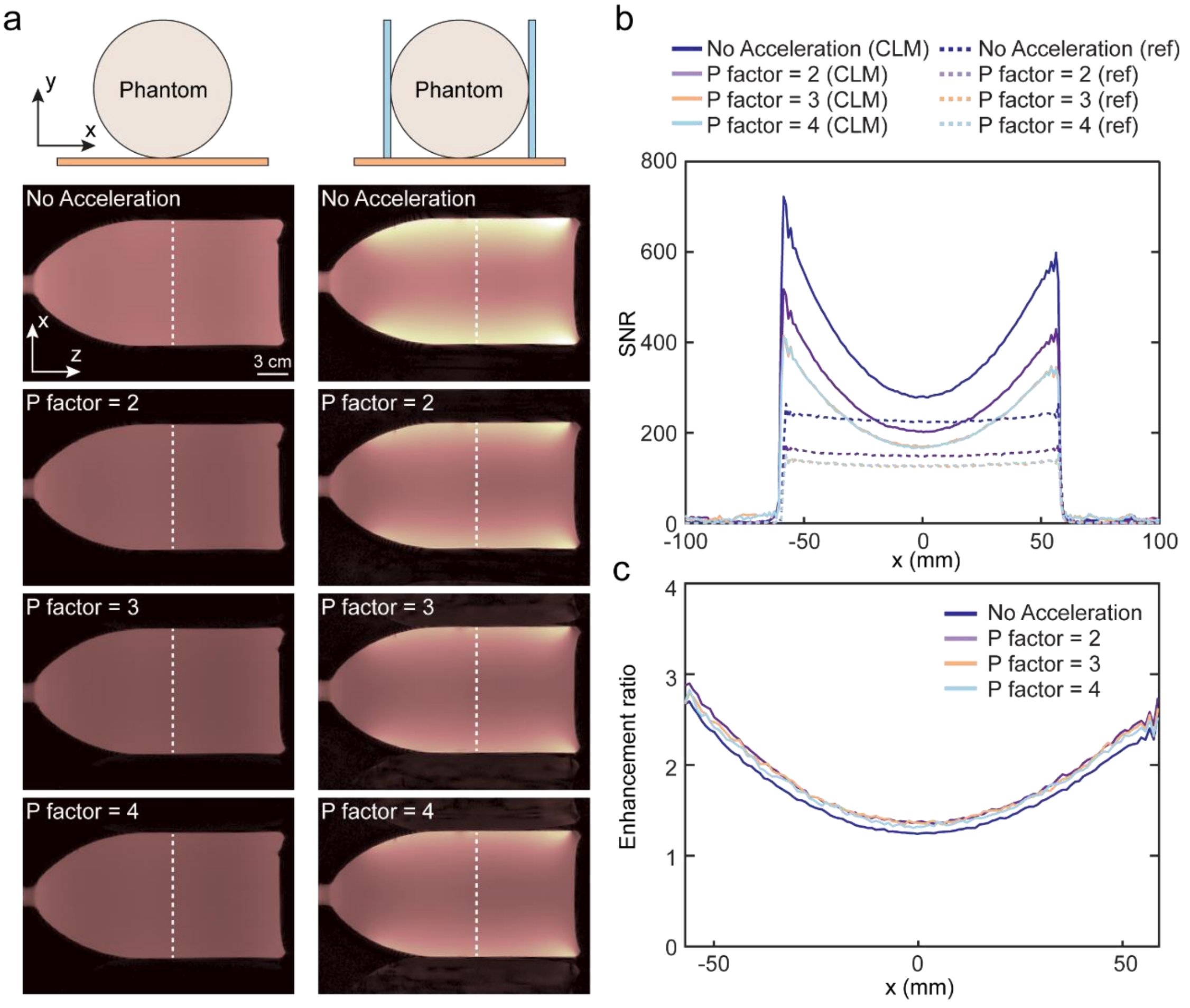

**Figure 5 | Compatibility of the CLM with parallel imaging acceleration.** a) Coronal SNR maps acquired without the CLM (Setup #1, reference) and with bilateral CLMs (Setup #4) under SENSE acceleration factors of P = 1 (no acceleration), 2, 3, and 4. b) SNR profiles extracted along the dashed lines in (a) and c) corresponding enhancement ratios between Setup #4 and Setup #1. Scale bars in (a) are 3cm.

## 2.6 Ex Vivo Validation

To evaluate the CLM's performance on biological tissue with heterogeneous internal structure, an ex vivo porcine tissue specimen was imaged using Setup #4 (bilateral CLMs with posterior coil) and Setup #1 (posterior coil only, reference). Three standard clinical pulse sequences were employed: T1-weighted turbo spin echo (T1w-TSE), T2-weighted turbo spin echo (T2w-TSE), and proton density-weighted turbo spin echo with fat saturation (PDw-TSE SPIR). These sequences exploit distinct tissue relaxation properties to generate complementary contrast mechanisms, and together they represent the core of routine musculoskeletal imaging protocols. By comparing the CLM-enhanced and reference images across all three weightings, the sequence-independence of the CLM enhancement can be assessed. **Figure 6a** presents the resulting coronal images for both configurations. In this imaging plane, the two bilateral CLMs are oriented along the upper and lower edges of the field of view, flanking the specimen from both sides. Across all three sequences, the CLM configuration produces visibly improved signal intensity, sharper delineation of internal anatomical features such as muscle fiber bundles and fascial boundaries, and reduced apparent noise granularity compared with the reference.

To quantify these improvements, three ROIs were selected within the T1w-TSE images at locations representing distinct spatial positions relative to the CLM (**Figure 6a**): ROI 1 and ROI 3 are positioned in the peripheral muscle regions adjacent to the bilateral CLMs, while ROI 2 is located in the deeper central region of the specimen. The SNR enhancement ratios for each ROI across all three sequences are shown in **Figure 6b**. ROIs 1 and 3 exhibit enhancement ratios of approximately 4-5-fold, consistent with their proximity to the metasurface where the near-field enhancement is strongest. ROI 2, located at the center of the specimen and farther from both CLMs, yields a lower but still substantial enhancement of approximately 2.5–3-fold. The enhancement ratios remain broadly consistent across the three pulse sequences, with only minor variations attributable to differences in baseline SNR and tissue-contrast-dependent signal weighting. This sequence-independence confirms that the CLM enhances the received RF signal at the electromagnetic level, prior to contrast encoding, and therefore does not distort the tissue-specific contrast generated by the pulse sequence.

Parallel imaging compatibility was further validated by acquiring coronal T1w-TSE images under SENSE acceleration factors of P = 2, 3, and 4 for both configurations (**Figure 6c**). The corresponding ROI-based enhancement ratios (**Figure 6d**) remain remarkably stable across all acceleration factors, with all three ROIs maintaining consistent enhancement values regardless of the degree of acceleration. This result corroborates the phantom-based findings of Section 2.5, confirming that the CLM preserves the spatial encoding diversity required for parallel imaging reconstruction even in the presence of structurally complex biological tissue. Notably, at P = 4, the reference images exhibit a pronounced residual aliasing artifact (red arrows, **Figure 6c**), whereas the corresponding CLM-enhanced images show substantially suppressed aliasing at the same location. This attenuation is a consequence of the elevated local SNR provided by the CLM: because the aliased signal component is superimposed on a higher-amplitude true signal, the relative contribution of the aliasing artifact to the image is diminished, resulting in improved visual fidelity of the reconstructed image under aggressive acceleration.

To verify that the CLM remains fully detuned during RF transmission under these imaging conditions, a $B_1^+$ map was acquired using a gradient-echo-based dual-TR method with the CLM in place (**Figure 6e**). The resulting map shows a spatially smooth transmit field distribution across the specimen, with no evidence of localized $B_1^+$ concentration or distortion in the vicinity of the CLMs. This confirms that the self-detuning mechanism effectively suppresses resonant interaction during the high-power transmit phase, consistent with the bench-level characterization presented in Section 2.2.

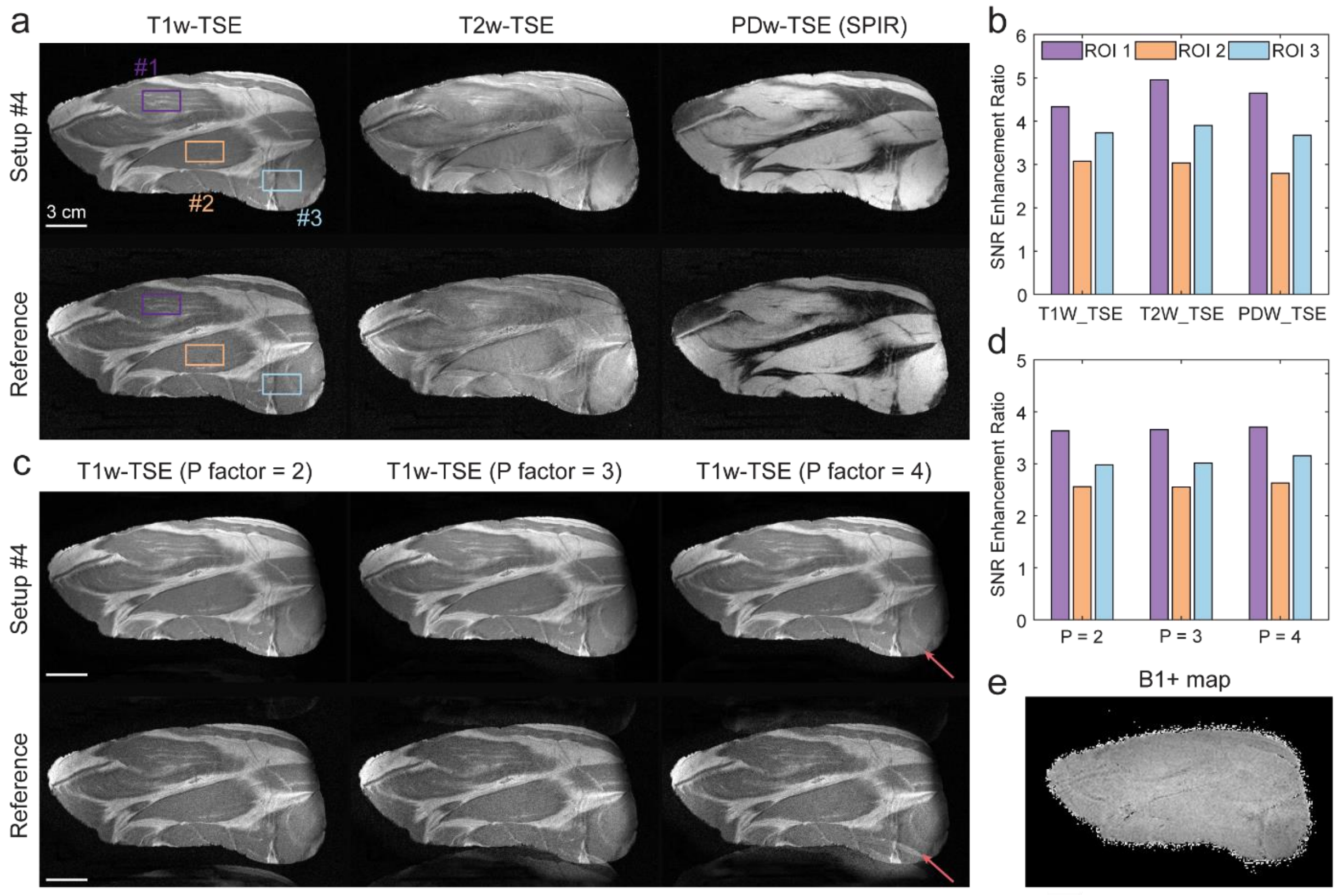

**Figure 6 | Ex vivo validation of the CLM across clinical sequences and accelerated imaging.** a) Coronal MR images of an ex vivo porcine specimen acquired with bilateral CLMs (Setup #4) and without CLM (reference) using T1w-TSE, T2w-TSE, and PDw-TSE (SPIR). b) ROI-based SNR enhancement ratios across sequences. c) Coronal T1w-TSE images under SENSE acceleration (P = 2-4). The CLM maintains a higher signal and improved image quality at increasing acceleration factors. d) Enhancement ratios for ROI #1, #2, and #3 under parallel imaging, confirming preserved enhancement performance. e) $B_1^+$ map of the CLM scanned using the dual-TR method, showing a smooth transmit field, indicating effective self-detuning. Scale bars in (a), (c), and (e) are 3cm.

### 2.7 In Vivo Validation

To further assess the performance of the CLM under realistic imaging conditions, in vivo imaging was conducted using the bilateral CLM configuration (Setup #4) in combination with the posterior coil, with the posterior coil-only setup serving as the reference. The subject was positioned supine on the patient bed, with two CLMs placed bilaterally on either side of the ankle. For the reference acquisition, the CLMs were removed while maintaining identical subject positioning and imaging conditions. All images were acquired using a T2-weighted turbo spin echo (T2w-TSE) sequence (Tables S3, Supporting Information). Representative sagittal and coronal images are shown in **Figure 7**. When used as an insert for the posterior coil, the CLM-enhanced images exhibit visibly increased signal intensity and improved delineation of anatomical structures within the joint region. The zoomed-in views further highlight the contrast and signal differences between the two configurations. Fine structural features, including cartilage boundaries and surrounding soft tissue interfaces, appear more clearly defined with the CLM in place. This enhancement is observed consistently across both imaging planes.

It is worth noting that the posterior coil used in this experiment is not specifically optimized for ankle imaging, and therefore does not represent the performance of dedicated multi-channel coils designed for this anatomical region. The purpose of this experiment is not to compete with such specialized hardware, but rather to demonstrate the role of the CLM as a flexible, insertable enhancement module that can be deployed in conjunction with existing receive arrays. In this configuration, the CLM provides localized signal enhancement in a place-and-play manner, without requiring any modifications to the scanner hardware or imaging protocol. This introduces a degree of operational flexibility, allowing the enhancement to be positioned according to the specific ROI, a capability particularly relevant in anatomical scenarios not readily served by fixed-geometry receive coils. Although the present example focuses on ankle imaging, the same principle is applicable to other anatomical regions, consistent with the flexible deployment strategies illustrated in **Figure 4a**. These results demonstrate that the CLM maintains its functionality in vivo and supports its potential as a reconfigurable, insertable solution for localized SNR enhancement.

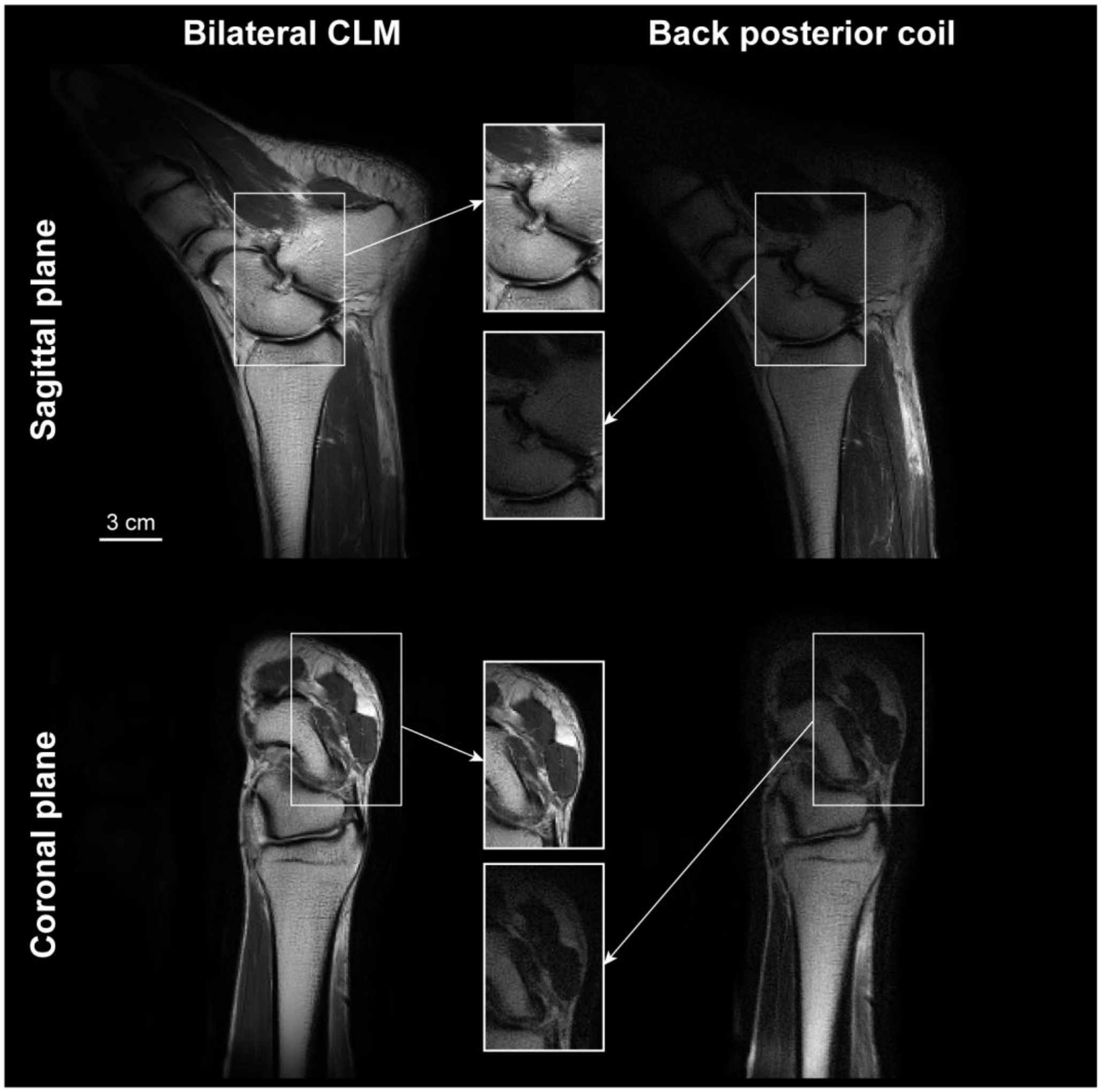

**Figure 7 | In vivo performance of the bilateral CLM in ankle imaging.** Representative sagittal and coronal T2w-TSE images of the ankle acquired using the posterior coil with and without the bilateral CLM. The CLM-enhanced images exhibit increased signal intensity and improved delineation of anatomical structures, as highlighted in the zoomed-in views.

## Discussion

In this work, we have introduced the CLM as a passive, wireless device for localized SNR enhancement in MRI. Through systematic electromagnetic characterization, phantom validation, and ex- and in-vivo imaging, we have demonstrated that the CLM achieves three capabilities in a single platform: reconfigurable field enhancement free of destructive interference, insertable deployment alongside standard clinical receive arrays, and full compatibility with parallel imaging acceleration. Such a combination of attributes, realized within a single platform, is seldom observed in previously reported metamaterial or metasurface designs for MRI.

The performance of the CLM is rooted in a physical mechanism that is fundamentally distinct from that of conventional arrayed metamaterials. In traditional designs, scaling up the number of resonant elements inevitably introduces inter-element coupling that drives opposing currents and creates localized signal voids. The CLM circumvents this limitation through its nested loop topology and shared current paths, which enforce in-phase current flow across all loops regardless of the array density. This collective behavior suppresses signal voids and is accompanied by a progressive increase in magnetic field strength with increasing loop count. The saturation of peak SNR enhancement observed beyond the 6-loop configuration reflects diminishing marginal coupling gains as rung densification continues within a fixed metasurface footprint. In practice, this does not constrain the utility of the CLM, as the enhancement area can be extended by increasing the physical dimensions of the metasurface rather than indefinitely increasing the rung density.

As with all near-field enhancement devices, the SNR gain provided by the CLM is spatially non-uniform along the normal direction and decays with distance from the metasurface. Rather than constituting a fundamental limitation, this characteristic aligns naturally with the insertable operational paradigm demonstrated in this work. By positioning the CLM in proximity to the ROI, whether beneath, beside, or above the imaging subject, the clinician can exploit the strong near-field enhancement precisely where it is most needed. The experiments with multiple CLM orientations relative to a clinical posterior coil illustrate that the enhancement direction can be tailored simply by repositioning the device, without any hardware or software reconfiguration. The consistent enhancement ratio observed under parallel imaging acceleration factors up to four-fold further confirms that this flexibility extends to accelerated acquisition protocols that dominate modern clinical practice.

From a translational perspective, the CLM is constructed entirely from commercially available components, including standard coaxial cables, conventional PCBs, and off-the-shelf PIN diodes, requiring no specialized fabrication processes. The material cost of a single CLM unit is substantially lower than that of a dedicated multi-channel receive coil, and its mechanical simplicity allows rapid assembly and reconfiguration by clinical staff without engineering expertise. Future development may explore flexible or conformal substrate designs that allow the CLM to conform to curved anatomical surfaces, as well as extension to other field strengths, since the underlying resonance mechanism of the CLM is not intrinsically limited to 3.0 T and can be retuned through capacitive adjustment.

## Methods

### Metamaterial Construction

The non-magnetic coaxial cable used in this study belonged to the RG174 cable group. The diameters of the inner conductor, dielectric insulation, outer conductor, and outer jacket were 0.48 mm, 1.42 mm, 1.93 mm, and 2.54 mm, respectively. The CLM coils were fabricated using coaxial cable segments as resonant elements. Each unit-cell rung consisted of two 105-mm-long coaxial cable segments separated by a 3-mm outer gap, with the outer conductors electrically connected by soldering. Multiple rungs were then soldered onto two shared printed circuit board (PCB) traces, forming the CLM structure. The PCBs were fabricated by milling 35-µm-thick copper cladding on a 1.5-mm-thick FR-4 substrate using a printed circuit board prototyping machine (ProtoMat S64, LPKF). PIN diodes were connected between the inner and outer conductors at both ends of each coaxial segment to enable the nonlinear behavior.

**Simulations**

All electromagnetic simulations of the magnetic-field (H-field), electric-field, and surface current distributions were performed using the frequency-domain solver in CST Microwave Studio Suite 2021. The geometrical dimensions used in the simulations matched those of the experimentally constructed CLM coils with different loop numbers. SAR distributions were calculated using the time-domain solver with the human voxel model Gustav from the CST voxel family. SAR values were calculated using the MRI toolbox in CST during post-processing and were normalized to 1 W of accepted power. PIN diodes were modeled as 1.2-pF capacitors in the receive state and as short-circuited paths in the transmit state, consistent with established approximations used in metamaterial MRI studies. To evaluate the interaction between the CLM and the RF transmit system, a high-pass birdcage coil was modeled to generate a circularly polarized RF field. The BC coil consisted of 16 legs with a length of 800 mm, 32 lumped capacitors with a capacitance of 10 pF, and two discrete excitation ports with a 90° phase shift to achieve quadrature excitation.

**Characterizations**

Electromagnetic (EM) characterization was performed using a vector network analyzer (VNA, E5071C, Keysight Inc.). An inductive loop probe was used to excite the coaxial coils during the measurements. The reflection coefficient S11 was measured while sweeping the excitation power from -12 to 10 dBm.

**Figure Preparation**

Three-dimensional models and device renderings were generated using SolidWorks 2021 and KeyShot 2024. Scientific illustrations, annotations, and figure layouts were prepared using Adobe Illustrator 2025. Data plots and image maps were generated using MATLAB R2021b. Electromagnetic simulations and simulation-based field maps were performed and exported using CST Studio Suite 2021.

**MRI Validation**

MRI experiments performed in Figures 3-5 used the mineral oil phantom, adopting the gradient echo pulse sequence for imaging. The mineral oil phantom has the following geometric parameters: diameter 130 mm, height 200 mm, relative permittivity 2.1, and conductivity 0.175 S $m^{-1}$. Parallel imaging validation was conducted with the SENSE reconstruction algorithm. The FA was set to 90° for all phantom experiments unless otherwise indicated. Detailed information about the experimental setup and sequence parameters of the MRI validations can be found in Figure S5, Table S2, and Table S3 (Supporting Information). The ex vivo porcine samples used in this study were obtained from a local abattoir.

**In Vivo Validation**

In vivo MRI validation was performed under an approved Institutional Review Board protocol at Boston University, with ethical approval obtained under IRB protocol H-45577. One healthy volunteer was included for proof-of-concept in vivo validation. The participant was positioned supine on the MRI patient table, and the ankle was imaged using the FlexCoverage Posterior coil array (Philips Healthcare) as the receive array, with the Birdcage coil used for RF transmission. For CLM-enhanced imaging, two CLMs were positioned bilaterally on either side of the ankle, corresponding to Setup #4 in Figure 4. For the reference acquisition, the CLMs were removed while maintaining the same participant positioning, coil configuration, imaging plane, and sequence parameters. Images were acquired on a 3.0 T clinical MRI system using a T2-weighted turbo spin echo (T2w-TSE) sequence. Detailed sequence parameters are provided in Table S3, Supporting Information. The acquired images were compared between the CLM-enhanced and reference configurations to evaluate localized signal enhancement and anatomical delineation. The participant was monitored during the scan, and no discomfort or adverse events were reported.

**Data availability**

The data that support the findings of this study are available from the corresponding author upon reasonable request.

**Acknowledgements**

This research was supported by the Rajen Kilachand Fund for Integrated Life Science and Engineering. The authors thank Andrew Ellison for valuable assistance and discussions during MRI experiments. The authors thank the Boston University Photonics Center for technical support.

**Author contribution**

Y. Liu and X. Zhu. contributed equally to this work. Y. Liu, X. Zhu, S. W. Anderson, and X. Zhang conceived the study. Y. Liu, X. Zhu, and X. Zhang designed and constructed the metamaterial. Y. Liu, X. Zhu and X. Zhang designed and conducted the bench measurements. Y. Liu, X. Zhu, K.

Wu, A. Kaliaev, C. LeBedis and X. Zhang conducted the MRI scans. All authors participated in discussing the results. Y. Liu, X. Zhu, S. W. Anderson, and X. Zhang drafted the manuscript. S. W. Anderson, and X. Zhang provided project supervision.

**Institutional Review Board Statement**

Prior to beginning the study, approval was obtained from our Institutional Review Board (IRB), with ethical approval obtained under protocol H-45577. Informed consent was obtained from the participant.

**Competing interests**

The authors have filed patent applications on the work described herein, application No.: 16/002,458, 16/443,126, and 17/065,812. Applicant: Trustees of Boston University. Inventors: Xin Zhang, Stephan Anderson, Guangwu Duan, and Xiaoguang Zhao. Status: Active.

**Additional information**

Additional information is available for this work.